\documentstyle[preprint,aps]{revtex}
\begin{document}
\baselineskip 0.7cm

{\bf T\"ornqvist and Roos reply:}
\bigskip

The comment of Harada et al. [1] supports our claim expressed in our
reply to Isgur and Speth [2] that crossed channel singularities are included
in an average sense through the contact term and form factor $F(s)$, and that
the $\sigma$ is needed to describe the data.
There are in fact three recent analyses by different groups [3-5], not
mentioned in Ref. [6] nor [7], which using different models 
find that a broad $\sigma$ with mass $\approx 500$ MeV is required by the data.
These together with the present comment strengthen the actual existence of the 
light and broad $\sigma$.

In this connection a few general remarks are in order. First, we do not find
it useful to study separately contributions from contact, crossed channel
and resonances (such as $\sigma$), since already by chiral symmetry [8] such
Born terms are intimately intertwined with each others. Each diagram 
can separately violate the symmetry, and e.g. the contact and resonance 
term interfere  destructively very 
strongly producing the Adler zeroes. Similar situations
apply for gauge symmetries. Furthermore, after unitarization  
any such term in a sum   of Born terms
becomes connected to all other terms in the chain of
higher order  multiloop diagrams. The resulting amplitude
 is then (c.f. the N/D formalism)
rather a {\it product} of $s$-channel resonances (in $D^{-1}(s)$) 
and crossed channel exchanges (in $N(s)$). In such a product a separation
of a broad resonance pole from the "background" 
always involves analytic continuation, which is sensitive to fine details.
Small changes in the amplitude at the real $s$-axis, where the data is,
can lead to a very different picture in terms of separate terms in the sum.

Within a limited energy region (as from threshold to 1.5 GeV, which covers
the region under discussion) there is very little freedom in chosing the 
$N(s)$ function, or its generalization in our multichannel framework:
$\propto (s-s_{Ai})F^2(s)$ (See especially Sects 2.6-7 of Ref. [9]).
Moreover, using flavor symmetry and data on $a_0(980)$ and $K^*_0(1430)$
this "freedom" is further reduced.
In fact, since our simple parametrization of $F(s)$  fits the data,  
one can argue that $F(s)$ is a rough
experimental representation of the crossed channel contributions.
In accord with this, one observes that the parameter $k_0$ in
$F^2(s)=exp[-k^2/k_0^2]$ is of the right order of magnitude,
expected from typical
meson sizes. Using the formula of LeYaouanc et al. [10] (neglecting 
polynomial factors) this gives a radius of $R=\sqrt 6/k_0=0.86$ fm.
Since our model does include crossed channel singularities in this approximate
sense, we do not support the suggestion  of Harada et al. [1],
that our $\sigma$ parameters will change much in a more refined model.

Of course, our conjecture that $(s-s_{Ai})F^2(s)$ is 
approximately of the correct form for the $N(s)$ functions
should be proven by a more elaborate analysis, including sufficiently 
many of the crossed channel singularities, in addition to the
already included chiral symmetry,
flavor symmetry, unitarity and analyticity
. 
Before the advent of
such an analysis, we must be content with cruder models, such as that of 
Harada et al. [1, 3], which supports, although does not prove, our conjecture. 

\bigskip

[1]  M. Harada, F. Sannino and J. Schechter, previous comment.

[2] N. Isgur and J. Speth, Phys. Rev. Lett. {\bf 77} 2332 (1996) and
N. A. T\"ornqvist and M. Roos, {\it ibid} 2333 (1996).

[3]  M. Harada, F. Sannino and J. Schechter, Phys. Rev.{\bf D54} 1991 (1996).

[4] S. Ishida et al., Prog. Theor. Phys. {\bf 95} 745  (1995), and
 KEK preprint 96-131 (Oct 1996), hep-th/9610359.

[5] R. Kami\'nski, L. Le\'sniak and J.-P. Maillet, Phys. Rev. {\bf D 50} 
3145 (1994) and R. Kami\'nski, L. Le\'sniak and K. Rybicki, Cracow preprint
No. 1730/Ph, hep-ph 9606362, (to appear in Z. Phys. C).
 
[6] N.A. T\"ornqvist and M. Roos, Phys. Rev. Lett. {\bf 76} 1575 (1996).

[7] R. M. Barnett et al., (the Particle Data Group),  Phys. Rev. {\bf D54} 1
(1996).

[8] De Alfaro et al., "Currents and Hadron Physics", North Holland (1973) p.
324-327.

[9] N.A. T\"ornqvist, Z. Phys. C {\bf68} 647 (1995).

[10] LeYaouanc et al., Phys. Rev {\bf D8 } 2223 (1973).

\bigskip
Nils A. T\"ornqvist$^{1)}$ and Matts Roos$^{2)}$
\bigskip

$^{1)}$ Research Institute for High Energy Physics, SEFT, POB 9, FIN-00014, 
Univer\-sity of Hel\-sinki, Fin\-land

$^{2)}$ High Energy Physics Laboratory, POB 9, FIN-00014, Univer\-sity of 
Hel\-sinki, Fin\-land

\bigskip\bigskip

Yours sincerely, 
\vskip 1cm
\hskip 6cm Nils T\"ornqvist, Matts Roos

\hskip 6cm Tornqvist@phcu.Helsinki.Fi
\vfill
\end{document}